\documentstyle[aps]{revtex}

\def\dt{{\delta \theta}}

\def\dt2{(\delta \theta)^2}

\def\de{\delta}

\def\nonum{\nonumber}

\def\dt{\frac{\partial}{\partial T}}

\def\jo #1#2#3#4{#1 {\bf #2}, #4  (#3)}
\def\PR{Phys.\ Rev.}
\def\PRB{Phys.\ Rev.\ B}
\def\PRL{Phys.\ Rev.\ Lett.}

\def\SSC{Solid State Commun.}

\def\JPF{J.\ Phys.\ France}

\begin{document}
\draft
\twocolumn[    
\widetext      
\title{Electronic Properties of Armchair Carbon Nanotubes : Bosonization Approach}
\author{Hideo Yoshioka$^{1,2}$ and Arkadi A. Odintsov$^{1,3}$}
\address{
$^{1}$Department of Applied Physics, Delft University of Technology, 
Lorentzweg 1, 2628 CJ Delft, The Netherlands. \\
$^{2}$Department of Physics, Nagoya University, 
Nagoya 464-01, Japan. \\
$^{3}$Nuclear Physics Institute, Moscow State University, 
Moscow 119899 GSP, Russia. \\
}
\date{\today}
\maketitle
\vspace{-1.0cm} 
\widetext                  
\begin{abstract} 
\leftskip 54.8pt           
\rightskip 54.8pt          

The phase Hamiltonian of armchair carbon nanotubes at half-filling and away
from it is derived from the microscopic lattice model by taking the long
range Coulomb interaction into account. We investigate the low energy
properties of the system using the renormalization group method. At
half-filling, the ground state is a Mott insulator with spin gap, in which
the bound states of electrons at different atomic sublattices are formed.
The difference from the recent results {[Phys. Rev. Lett. {\bf 79}, 5082
(1997)]} away half-filling is clarified.
\end{abstract}


]                

\narrowtext
Single wall carbon nanotubes (SWNTs) with diameters of a few atomic
distances and lengths of several micrometers can be considered as the
ultimate miniaturization of metallic wires \cite{Thess}. Recent experiments
have demonstrated electron transport through individual \cite{Tans} and
multiple \cite{Bockrath} SWNTs as well as provide evidence of strong Coulomb
interaction in these systems. The one-dimensional nature of the low-energy
electronic states in the nanotubes together with the interaction of
electrons should result in a variety of correlation effects due to the
non-Fermi liquid ground state of the system \cite{Voit}.

Very recent transport spectroscopy data by Tans et. al. \cite{Tans2} on spin
polarization of an individual SWNT can not be explained by the constant
interaction model and suggests the interpretation in terms of electron
correlations. This result, however, was not confirmed by experiments on
ropes of SWNTs \cite{Cobden}, which fit the constant interaction model
remarkably well.

Experimental progress urges the development of a theory of electron
correlations in SWNTs. For a model on-site \cite{Balents-Fisher} and on-site
plus nearest neighbor \cite{Krotov-Lee-Louie} interactions, metallic
armchair SWNTs become Mott insulator at half-filling, whereas upon doping
they exhibit superconducting fluctuations. 
The realistic long-range Coulomb interaction
was considered in Refs. \cite{Egger-Gogolin,Kane-Balents-Fisher}.  
Kane, Balents and Fisher \cite{Kane-Balents-Fisher} discussed the
effects of the Coulomb interaction in finite size
armchair nanotubes (ANs) in terms of the Tomonaga-Luttinger low energy
theory.
The most important part of the forward scattering 
was incorporated into the Tomonaga-Luttinger-like Hamiltonian,
whereas the other types of scattering were 
treated as perturbations.
Egger and Gogolin (EG) developed an
effective low energy theory of ANs starting from a microscopic model \cite
{Egger-Gogolin}, which accounts for all types of scattering processes. They
derived a bosonic phase Hamiltonian and discussed possible ground state away
from half-filling.

In this Letter we derive the phase Hamiltonian of ANs and evaluate its
parameters from the microscopic lattice model. The difference between our
Hamiltonian and that by EG stems from the distinction in the form of kinetic
term and the use of oversimplified approximation for $2k_{F}$-component of
scattering amplitudes in Ref. \cite{Egger-Gogolin}, which overlooks $C_{3}$
symmetry of the microscopic lattice model. The renormalization group (RG)
method is applied to the Hamiltonian, and the low energy states are
investigated. At half-filling the ground state is the Mott insulator with
spin gap, in agreement with the conclusion of Hubbard-like models \cite
{Balents-Fisher,Krotov-Lee-Louie}. In this state, the electronic bound
states are formed between the different sublattices. Away from half-filling
we predict gaps for both symmetric and antisymmetric spin modes, in contrast
to the result of Ref. \cite{Egger-Gogolin} for the case of equal amplitudes
of intrasublattice and intersublattice forward scattering. 

\smallskip We start from the standard tight-binding single particle
Hamiltonian \cite{Wallace} for $p_{z}$ electrons on the honeycomb lattice
(inset of Fig. 1), 
\begin{equation}
{\cal H}_{k}=\sum_{s,\vec{k}}\left\{ \xi (\vec{k})a_{-,s}^{\dagger }(\vec{k}%
)a_{+,s}(\vec{k})+h.c.\right\} .  \label{eqn:1}
\end{equation}
Here $a_{p,s}(\vec{k})$ are the Fermi operators for electrons at the
sublattice $p=\pm $ with the spin $s=\pm $ and the wavevector $\vec{k}%
=(k_{x},k_{y})$. The matrix elements are given by $\xi (\vec{k})=-t({\rm e}%
^{-{\rm i}k_{y}a/\sqrt{3}}+2{\rm e}^{{\rm i}k_{y}a/2\sqrt{3}}\cos k_{x}a/2)$%
, $t$ being the hopping amplitude between neighboring atoms. The eigenvalues
of the Hamiltonian vanish at two points of the Brillouin zone, $\vec{k}%
=(\alpha K_{0},0)$ with $\alpha =\pm $ and $K_{0}=4\pi /3a$, which
constitute the Fermi surface of a graphite layer \cite
{Wallace,Balents-Fisher}.

\smallskip We consider the armchair $(N,N)$ SWNT parallel to the $x$ axis so
that the wrapping vector $\vec{w}=N(\vec{a}_{+}+\vec{a}_{-})$ points in $y$
direction (inset of Fig.1). In this case the Fermi points lie on the allowed
quantized transverse wavevector $k_{y}=0$ for any $N$. Expanding Eq. (\ref
{eqn:1}) near the Fermi points to the lowest order in $q=k_{x}-\alpha K_{0}$
and introducing slowly varying Fermi fields $\psi _{p\alpha
s}(x)=L^{-1/2}\sum_{q}e^{iqx}a_{p,s}(q+\alpha K_{0},0)$, we obtain 
\begin{equation}
{\cal H}_{k}=-iv_{0}\sum_{p,\alpha ,s}\alpha \int {\rm d}x\psi _{-p\alpha
s}^{\dagger }\partial _{x}\psi _{p\alpha s},  \label{Hkpsi}
\end{equation}
$v_{0}=\sqrt{3}ta/2\approx 8\times 10^{5}$ m/s being the Fermi velocity. It
should be noted \cite{difference} that Eq. (\ref{Hkpsi}) differs from Eq.
(2) of Ref. \cite{Egger-Gogolin}. Despite identical energy spectra of both
Hamiltonians, the phase relations between the components of an eigenfunction
at the two sublattices are different. The consequences of this fact will be
discussed later on.

Following EG, the interaction term of the Hamiltonian reads, 
\begin{eqnarray}
{\cal H}_{int} &=&\frac{1}{2}\left( \frac{a}{2}\right) ^{2}\sum_{l,l^{\prime
}}\sum_{pp^{\prime }}\sum_{\alpha _{1}...\alpha _{4}}\sum_{ss^{\prime
}}U_{pp^{\prime }}(x_{l}-x_{l^{\prime }})  \label{Hint} \\
&&\times \psi _{p\alpha _{1}s}^{\dagger }(x_{l})\psi _{p^{\prime }\alpha
_{2}s^{\prime }}^{\dagger }(x_{l^{\prime }})\psi _{p^{\prime }\alpha
_{3}s^{\prime }}(x_{l^{\prime }})\psi _{p\alpha _{4}s}(x_{l}),  \nonumber
\end{eqnarray}
with $x_{l}=la/2$. The effective 1D interaction between the sublattices $p$
and $p^{\prime }$, $U_{pp^{\prime }}(x_{l})$, is the average of the Coulomb
potential $U(x,y)={e^{2}}/\{\kappa \sqrt{a_{0}^{2}+x^{2}+4R^{2}\sin
^{2}(y/2R)}\}$ over the nodes of a sublattice along the $y$ direction, 
\begin{equation}
U_{pp^{\prime }}\left( x_{l}\right) =\frac{1}{N}\sum_{n}U\left( x_{l},a\sqrt{%
3}(n+\Delta _{pp^{\prime }})\right) ,  \label{Vpp'}
\end{equation}
with $\Delta _{pp^{\prime }}=%
\mathop{\rm mod}%
(l,2)/2+\delta _{pp^{\prime }}/3$. Here $\kappa $ is an effective dielectric
constant of the system (the estimate \cite{Egger-Gogolin} for the parameters
of the experiment \cite{Tans} gives $\kappa =1.4$) and $a_{0}\simeq a$
characterizes the radius of $p_{z}$ orbital wave function in the graphite
plane.

The Hamiltonian ${\cal H}_{int}$ can be separated into the ''forward
scattering'' ${\cal H}_{0}+{\cal H}_{f}$ ($\alpha _{1}=\alpha _{4}$, $\alpha
_{2}=\alpha _{3}$) and ''backscattering'' ${\cal H}_{b}+{\cal H}_{b^{\prime
}}$ ($\alpha _{1}=-\alpha _{2}=\alpha _{3}=-\alpha _{4}$) \cite{commentfbs}, 
\begin{eqnarray}
{\cal H}_{0} &=&\frac{V_{pp}(0)}{2}\int {\rm d}x\rho ^{2}(x),  \label{H0} \\
{\cal H}_{f} &=&-\frac{\delta V(0)}{2}\sum_{p\alpha \alpha ^{\prime
}ss^{\prime }}\int {\rm d}x\psi _{p\alpha s}^{\dagger }\psi _{-p\alpha
^{\prime }s^{\prime }}^{\dagger }\psi _{-p\alpha ^{\prime }s^{\prime }}\psi
_{p\alpha s},  \label{Hf} \\
{\cal H}_{b} &=&\frac{V_{pp}(2K_{0})}{2}\sum_{pp^{\prime }\alpha ss^{\prime
}}\int {\rm d}x\psi _{p\alpha s}^{\dagger }\psi _{p^{\prime }-\alpha
s^{\prime }}^{\dagger }\psi _{p^{\prime }\alpha s^{\prime }}\psi _{p-\alpha
s},  \label{Hb} \\
{\cal H}_{b^{\prime }} &=&-\frac{\delta V(2K_{0})}{2}\sum_{p\alpha
ss^{\prime }}\int {\rm d}x\psi _{p\alpha s}^{\dagger }\psi _{-p-\alpha
s^{\prime }}^{\dagger }\psi _{-p\alpha s^{\prime }}\psi _{p-\alpha s},
\label{HbD}
\end{eqnarray}
where $\rho =\sum_{p\alpha s}\psi _{p\alpha s}^{\dagger }\psi _{p\alpha s}$
is the total electron density, $\delta V(0)={V}_{pp}(0)-{V}_{p-p}(0)$, and $%
\delta V(2K_{0})={V}_{pp}(2K_{0})-{V}_{p-p}(2K_{0})$, with ${V}_{pp^{\prime
}}(q)=(a/2)\sum_{l}e^{iqx_{l}}U_{pp^{\prime }}(x_{l})$ being the Fourier
transformed interaction.

The forward scattering (\ref{H0}) has the strongest amplitude, $%
V_{pp}(0)=2e^{2}/\kappa \ln (R_{s}/R)$, where $R_{s}\simeq \min (L,D)$
characterizes the screening of the Coulomb interaction due to a finite
length $L$ of the AN and/or the presence of metallic electrodes at a
distance $D$ \cite{Kane-Balents-Fisher}. The amplitude $V_{pp}(0)$ is
relatively insensitive to the choice of $R_{s}$ due to logarithmic
dependence. From Eq. (\ref{Vpp'}), one sees that the amplitudes $\delta V(0)$
and ${V}_{pp}(2K_{0})$ decay as $1/R$ for $R\gg a$. It should be noted that
the matrix element $N{V}_{p-p}(2K_{0})$ vanishes identically in case of a
graphite plane ($R\rightarrow \infty $) due to the $C_{3}$ symmetry of the
lattice. For this reason, ${V}_{p-p}(2K_{0})$ is much smaller than $\delta
V(0)$ and ${V}_{pp}(2K_{0})$. All the matrix elements decrease with
increasing $a_{0}$. Numerical evaluation for $a_{0}=a/2$ and $R\gg a$ gives $%
\delta V(0)=0.21$, ${V}_{pp}(2K_{0})=0.60$, ${V}_{p-p}(2K_{0})=9.4\times
10^{-4}$ in units of $ae^{2}/2\pi \kappa R$ (${V}_{p-p}(2K_{0})$ is
estimated for $N=10$). This result shows that the approximation ${V}%
_{pp}(2K_{0})={V}_{p-p}(2K_{0})$ used in Ref. \cite{Egger-Gogolin} is
questionable.

In order to bosonize the Hamiltonian ${\cal H}={\cal H}_{k}+{\cal H}_{int}$,
Eqs. (\ref{Hkpsi}), (\ref{H0})-(\ref{HbD}), we diagonalize the kinetic term
Eq. (\ref{Hkpsi}) by the unitary transformation 
\begin{equation}
\psi _{r\alpha s}=(\psi _{+\alpha s}+\alpha r\psi _{-\alpha s})/\sqrt{2},
\label{Utrans}
\end{equation}
which maps the basis of atomic sublattices ($p=\pm $) to the basis of right-
and left-movers ($r=\pm $). This transformation is different from one by EG
due to the different form of the kinetic term.

We bosonize the Fermi fields $\psi _{r\alpha s}$, 
\begin{equation}
\psi _{r\alpha s}=\frac{\eta _{r,\alpha ,s}}{\sqrt{2\pi a}}\exp \left[ {\rm i%
}rq_{F}x+\frac{{\rm i}r}{2}\left\{ \theta _{\alpha s}+r\phi _{\alpha
s}\right\} \right] ,   \label{psiras}
\end{equation}
and decompose the phase variables $\theta _{\alpha s},\phi _{\alpha s}$ into
symmetric $\delta =+$ and antisymmetric $\delta =-$ modes of the charge $%
\rho $ and spin $\sigma $ excitations, $\theta _{\alpha s}=\theta _{\rho
+}+s\theta _{\sigma +}+\alpha \theta _{\rho -}+\alpha s\theta _{\sigma -}$
and $\phi _{\alpha s}=\phi _{\rho +}+s\phi _{\sigma +}+\alpha \phi _{\rho
-}+\alpha s\phi _{\sigma -}$. The bosonic fields satisfy the commutation
relation, $[\theta _{j\delta }(x),\phi _{j^{\prime }\delta ^{\prime
}}(x^{\prime })]={\rm i}(\pi /2){\rm sign}(x-x^{\prime })\delta _{jj^{\prime
}}\delta _{\delta \delta ^{\prime }}$. The Majorana fermions $\eta _{r\alpha
s}$ are introduced to ensure correct anticommutation rules for different
species $r,\alpha ,s$ of electrons, and satisfy $[\eta _{r\alpha s},\eta
_{r^{\prime }\alpha ^{\prime }s^{\prime }}]_{+}=2\delta _{rr^{\prime
}}\delta _{\alpha \alpha ^{\prime }}\delta _{ss^{\prime }}$. The
spin-conserving products $\eta _{r\alpha s}\eta _{r^{\prime }\alpha ^{\prime
}s}$ in the Hamiltonian ${\cal H}$ can be represented as \cite{Egger-Gogolin}%
, $A_{++}(r,\alpha ,s)=\eta _{r\alpha s}\eta _{r\alpha s}=1$, $%
A_{+-}(r,\alpha ,s)=\eta _{r\alpha s}\eta _{r-\alpha s}={\rm i}\alpha \sigma
_{x}$, $A_{-+}(r,\alpha ,s)=\eta _{r\alpha s}\eta _{-r\alpha s}={\rm i}%
r\alpha \sigma _{z}$, and $A_{--}(r,\alpha ,s)=\eta _{r\alpha s}\eta
_{-r-\alpha s}=-{\rm i}r\sigma _{y}$ with the standard Pauli matrices $%
\sigma _{i}$ ($i=x,y,z$). The quantity $q_{F}=\pi n/4$ is related to the
deviation $n$ of the average electron density from half-filling, and can be
controlled by the gate voltage.

The bosonized Hamiltonian has the form, 
\begin{eqnarray}
{\cal H} &=&\sum_{j=\rho ,\sigma }\sum_{\delta =\pm }\frac{v_{j\delta }}{%
2\pi }\int {\rm d}x\left\{ K_{j\delta }^{-1}(\partial _{x}\theta _{j\delta
})^{2}+K_{j\delta }(\partial _{x}\phi _{j\delta })^{2}\right\}  \nonumber \\
&&+\frac{1}{2(\pi a)^{2}}\int {\rm d}x\{  \nonumber \\
&&[\delta V(0)-2\bar{V}(2K_{0})]\cos (4q_{F}x+2\theta _{\rho +})\cos 2\theta
_{\sigma +}  \nonumber \\
&&-\delta V(0)\cos (4q_{F}x+2\theta _{\rho +})\cos 2\theta _{\rho -} 
\nonumber \\
&&+\delta V(0)\cos (4q_{F}x+2\theta _{\rho +})\cos 2\theta _{\sigma -} 
\nonumber \\
&&-[\delta V(0)-\delta V(2K_{0})]\cos 2\theta _{\rho -}\cos 2\theta _{\sigma
-}  \nonumber \\
&&+{\delta V(0)}\cos 2\theta _{\sigma +}\cos 2\theta _{\sigma -}  \nonumber
\\
&&-{\delta V(0)}\cos 2\theta _{\sigma +}\cos 2\theta _{\rho -}  \nonumber \\
&&-{2\bar{V}(2K_{0})}\cos (4q_{F}x+2\theta _{\rho +})\cos 2\phi _{\sigma -} 
\nonumber \\
&&+{2\bar{V}(2K_{0})}\cos 2\theta _{\sigma +}\cos 2\phi _{\sigma -} 
\nonumber \\
&&+{\delta V(2K_{0})}\cos 2\theta _{\rho -}\cos 2\phi _{\sigma -}  \nonumber
\\
&&+{\delta V(2K_{0})}\cos 2\theta _{\sigma -}\cos 2\phi _{\sigma -}\},
\label{Hbos}
\end{eqnarray}
$v_{j\delta }=v_{0}\sqrt{A_{j\delta }B_{j\delta }}$ and $K_{j\delta }=\sqrt{%
B_{j\delta }/A_{j\delta }}$ being the velocities of excitations and
exponents for the modes $j,\delta $. The parameters $A_{j\delta }$, $%
B_{j\delta }$ are given by%
\begin{eqnarray}
A_{\rho +} &=&1+\frac{4\bar{V}(0)}{\pi v_{0}}-\frac{\delta V(0)}{4\pi v_{0}}-%
\frac{\bar{V}(2K_{0})}{2\pi v_{0}}-\frac{\delta {V}(2K_{0})}{4\pi v_{0}}\;\;,
\label{Arp} \\
B_{\rho +} &=&B_{\sigma +}=1+\frac{\delta V(0)}{4\pi v_{0}}+\frac{\bar{V}%
(2K_{0})}{2\pi v_{0}}-\frac{\delta {V}(2K_{0})}{4\pi v_{0}}\;\;,  \label{Brp}
\\
A_{\sigma +} &=&1-\frac{\delta V(0)}{4\pi v_{0}}-\frac{\bar{V}(2K_{0})}{2\pi
v_{0}}-\frac{\delta {V}(2K_{0})}{4\pi v_{0}}\;\;,  \label{Asp} \\
A_{\rho -} &=&A_{\sigma -}=1-\frac{\delta V(0)}{4\pi v_{0}}+\frac{\bar{V}%
(2K_{0})}{2\pi v_{0}}+\frac{\delta {V}(2K_{0})}{4\pi v_{0}}\;\;,  \label{Arm}
\\
B_{\rho -} &=&B_{\sigma -}=1+\frac{\delta V(0)}{4\pi v_{0}}-\frac{\bar{V}%
(2K_{0})}{2\pi v_{0}}+\frac{\delta {V}(2K_{0})}{4\pi v_{0}}\;\;,  \label{Brm}
\end{eqnarray}
with $\bar{V}(q)=[V_{pp}(q)+V_{p-p}(q)]/2$. The sublattice-independent
forward scattering $\bar{V}(0)$ strongly renormalizes the exponent for the
symmetric charge mode, $K_{\rho +}\approx 0.2$ \cite{Kane-Balents-Fisher},
whereas for the other modes the interaction is weak, $K_{j\delta }=1+O(a/R)$ 
\cite{Egger-Gogolin}.

We now compare the Hamiltonian (\ref{Hbos}) with that derived by EG \cite
{Egger-Gogolin} away from half-filling. In this case, the non-linear terms
containing the misfit parameter $q_{F}$ can be neglected due to the
breakdown of the momentum conservation. Despite the equal forward scattering
parts of both Hamiltonians, there is difference in the backscattering parts.
Namely, the Hamiltonian by EG can be obtained from ours by substituting ${%
\bar{V}}(2K_{0})\to 0$ and $\delta V(2K_{0})\to 2{\bar{V}}(2K_{0})$. The
absence of $\cos 2\theta _{\sigma +}\cos 2\phi _{\sigma -}$ term in the
Hamiltonian by EG originates from the use of the approximation $\delta
V(2K_{0})=0$, whereas the difference in the coefficient in front of the
three other non-linear backscattering terms stems from the difference in the
unitary transformation (\ref{Utrans}).

The low energy properties of the Hamiltonian Eq. (\ref{Hbos}) can be
investigated by the renormalization group (RG) method. The RG equations can
be derived by assuming scale invariance of the correlation functions $%
\left\langle \exp \left\{ {\rm i}\left( \theta _{j\delta }(x_{1},\tau
_{1})-\theta _{j\delta }(x_{2},\tau _{2})\right) \right\} \right\rangle $ 
\cite{Giamarchi-Schulz}. At half-filling, $q_{F}=0$, we obtain 
\begin{eqnarray}
(K_{\rho +})^{\prime } &=&-({K_{\rho +}^{2}}/{8}%
)(y_{1}^{2}+y_{2}^{2}+y_{3}^{2}+y_{7}^{2})\;\;,  \label{RG1} \\
(K_{\sigma +})^{\prime } &=&-({K_{\sigma +}^{2}}/{8}%
)(y_{1}^{2}+y_{5}^{2}+y_{6}^{2}+y_{8}^{2})\;\;,  \label{RG2} \\
(K_{\rho -})^{\prime } &=&-({K_{\rho -}^{2}}/{8}%
)(y_{2}^{2}+y_{4}^{2}+y_{6}^{2}+y_{9}^{2})\;\;,  \label{RG3} \\
(K_{\sigma -})^{\prime } &=&-({K_{\sigma -}^{2}}/{8}%
)(y_{3}^{2}+y_{4}^{2}+y_{5}^{2})  \nonumber \\
&+&({1}/{8})(y_{7}^{2}+y_{8}^{2}+y_{9}^{2})\;\;,  \label{RG4} \\
(y_{1})^{\prime } &=&\left\{ 2-(K_{\rho +}+K_{\sigma +})\right\} y_{1} 
\nonumber \\
&-&(y_{2}y_{6}+y_{3}y_{5}+y_{7}y_{8})/4\;\;,  \label{RG5} \\
(y_{2})^{\prime } &=&\left\{ 2-(K_{\rho +}+K_{\rho -})\right\} y_{2} 
\nonumber \\
&-&(y_{1}y_{6}+y_{3}y_{4}+y_{7}y_{9})/4\;\;,  \label{RG6} \\
(y_{3})^{\prime } &=&\left\{ 2-(K_{\rho +}+K_{\sigma -})\right\} y_{3} 
\nonumber \\
&-&(y_{1}y_{5}+y_{2}y_{4})/4\;\;,  \label{RG7} \\
(y_{4})^{\prime } &=&\left\{ 2-(K_{\rho -}+K_{\sigma -})\right\} y_{4} 
\nonumber \\
&-&(y_{2}y_{3}+y_{5}y_{6})/4\;\;,  \label{RG8} \\
(y_{5})^{\prime } &=&\left\{ 2-(K_{\sigma +}+K_{\sigma -})\right\} y_{5} 
\nonumber \\
&-&(y_{1}y_{3}+y_{4}y_{6})/4\;\;,  \label{RG9} \\
(y_{6})^{\prime } &=&\left\{ 2-(K_{\sigma +}+K_{\rho -})\right\} y_{6} 
\nonumber \\
&-&(y_{1}y_{2}+y_{4}y_{5}+y_{8}y_{9})/4\;\;,  \label{RG10} \\
(y_{7})^{\prime } &=&\left\{ 2-(K_{\rho +}+1/K_{\sigma -})\right\} y_{7} 
\nonumber \\
&-&(y_{1}y_{8}+y_{2}y_{9})/4\;\;,  \label{RG11} \\
(y_{8})^{\prime } &=&\left\{ 2-(K_{\sigma +}+1/K_{\sigma -})\right\} y_{8} 
\nonumber \\
&-&(y_{1}y_{7}+y_{6}y_{9})/4\;\;,  \label{RG12} \\
(y_{9})^{\prime } &=&\left\{ 2-(K_{\rho -}+1/K_{\sigma -})\right\} y_{9} 
\nonumber \\
&-&(y_{2}y_{7}+y_{6}y_{8})/4\;\;,  \label{RG13}
\end{eqnarray}
where $()^{\prime }$ denotes ${\rm d}/{\rm d}\ell $ with ${\rm d}\ell ={\rm d%
}\ln (\tilde{a}/a)$ ($\,\tilde{a}$ is the new lattice constant). The initial
conditions for the Eqs. (\ref{RG1})-(\ref{RG13}) are $K_{j\delta
}(0)=K_{j\delta }$, $y_{1}=[\delta V(0)-2\bar{V}(2K_{0})]/(\pi v_{0})$, $%
y_{2}=-y_{3}=-y_{5}=y_{6}=-\delta V(0)/(\pi v_{0})$, $y_{4}=-[\delta
V(0)-\delta V(2K_{0})]/(\pi v_{0})$, $y_{7}=-y_{8}=-2\bar{V}(2K_{0})/(\pi
v_{0})$, and $y_{9}=\delta V(2K_{0})/(\pi v_{0})$. In deriving the RG
equations, the non-linear term $\cos 2\theta _{\sigma -}\cos 2\phi _{\sigma
-}$ is omitted because this operator stays exactly marginal in all orders
and is thus decoupled from the problem \cite{Egger-Gogolin-2}. The RG
equations away from half-filling can be obtained from Eqs. (\ref{RG1})-(\ref
{RG13}) by putting $y_{1}$, $y_{2}$, $y_{3}$ and $y_{7}$ to zero. Hereafter
we concentrate on the case $N=10$, $\kappa =1.4$, $R_{s}$ = 100 $nm$ and $%
a_{0}=a/2$ where the initial values of the RG parameters correspond to the
estimates given below Eq. (\ref{HbD}).

\smallskip Away from half-filling, the quantities $K_{\sigma +}$, $K_{\rho -}
$ and $K_{\sigma -}^{-1}$ renormalize to zero and the coefficient of $\cos
2\theta _{\sigma +}\cos 2\theta _{\rho -}$($\cos 2\theta _{\sigma +}\cos
2\phi _{\sigma -}$and $\cos 2\theta _{\rho -}\cos 2\phi _{\sigma -}$) tends
to $-\infty $ ($\infty $). As a result, the phases $\theta _{\sigma
+},\theta _{\rho -}$ and $\phi _{\sigma -}$ are locked at $(\theta _{\sigma
+},\theta _{\rho -},\phi _{\sigma -})=(0,0,\pi /2)$ or $(\pi /2,\pi /2,0)$
so that the modes $\sigma \pm $ and $\rho -$ are gapped. In this case, the
asymptotic behavior of the correlation functions at $x\rightarrow \infty $
is determined by the correlations of the gapless $\rho +$ mode, $%
\left\langle {\rm e}^{{\rm i}n\theta _{\rho +}(x)}{\rm e}^{-{\rm i}n\theta
_{\rho +}(0)}\right\rangle \sim x^{-n^{2}K_{\rho +}/2}$ and $\left\langle 
{\rm e}^{{\rm i}m\phi _{\rho +}(x)}{\rm e}^{-{\rm i}m\phi _{\rho
+}(0)}\right\rangle \sim x^{-m^{2}/2K_{\rho +}}$ ($n=1$ and $2$ correspond
to $2q_{F}$ and $4q_{F}$ density waves and $m=1$ for a superconducting
state). Since $K_{\rho +}\approx 0.2$, the $2q_{F}$ density wave
correlations seem to be dominant. However, we found that the correlation
functions of any $2q_{F}$ density wave decay exponentially at large
distances due to the gapped modes. We therefore are looking for the
four-particle correlations. The $4q_{F}$ density waves dominate over the
superconductivity for $K_{\rho +}<1/2$ \cite{Nagaosa,Schulz}. Such density
wave states are given by the product of the charge $n_{\pm }(x)$ or spin $%
S_{\pm }(x)$ densities at different sublattices, 
\begin{eqnarray}
n_{+}(x)n_{-}(x) &\sim &-\frac{1}{2(\pi a)^{2}}\cos (4q_{F}x+2\theta _{\rho
+})  \nonumber \\
&\times &(2\cos 2\theta _{\sigma +}+\cos 2\phi _{\sigma -}-\cos 2\theta
_{\rho -}),  \label{nn} \\
S_{+}(x)S_{-}(x) &\sim &-\frac{1}{8(\pi a)^{2}}\cos (4q_{F}x+2\theta _{\rho
+})  \nonumber \\
&\times &(2\cos 2\theta _{\sigma +}-\cos 2\phi _{\sigma -}+\cos 2\theta
_{\rho -}),  \label{ss}
\end{eqnarray}
where we neglected the unlocked phases $\phi _{\sigma +},\phi _{\rho
-},\theta _{\sigma -}$ whose contribution decays exponentially at large
distances. Substituting the values of the locked phases we observe that $%
n_{+}(x)n_{-}(x)$ vanishes, and the dominant state is the $4q_{F}$ spin
density wave with correlation function $\left\langle
S_{+}(x)S_{-}(x)S_{+}(0)S_{-}(0)\right\rangle \sim \cos 4q_{F}x/x^{2K_{\rho
+}}$. 

The modes $\sigma \pm $ and $\rho -$ remain gapped also in the limit $\delta
V(0)=0$. In this case EG have obtained that the symmetric modes, $\rho +$
and $\sigma +$, are gapless, whereas the $\rho -$ mode is gapped and $\sigma
-$ mode separates into the gapless and gapped parts. The result by EG
follows from the special dual symmetry $\theta _{\sigma -}\leftrightarrow
\phi _{\sigma -}$ of the Hamiltonian and the absence of non-linear terms in
the $\rho +$ and $\sigma +$ sectors. Both these factors are lacking in Eq. (%
\ref{Hbos}). On the other hand, the result by EG for a finite value of $%
\delta V(0)$ is qualitatively the same as ours.

At half-filling the solution (Fig.1) of the RG equations (\ref{RG1})-(\ref
{RG13}) indicates that the phase variables $\theta _{\rho +}$, $\theta
_{\sigma +}$, $\theta _{\rho -}$, and $\phi _{\sigma -}$ are locked and the
all kinds of excitation are gapped. In other words, the ground state of the
half-filled AN is a Mott insulator with spin gap. The same conclusion has
been drawn from the model with short range interactions \cite
{Balents-Fisher,Krotov-Lee-Louie}. The locked phases are given by $(\theta
_{\rho +},\theta _{\sigma +},\theta _{\rho -},\phi _{\sigma -})=(0,0,0,0)$
or $(\pi /2,\pi /2,\pi /2,\pi /2)$ since the coefficients tend to $-\infty $
for the first, second and 6-9-th non-linear terms in Eq. (\ref{Hbos}). The
averages $\langle n_{+}(x)n_{-}(x)\rangle $ and $\langle
S_{+}(x)S_{-}(x)\rangle $ are both finite, which indicates the formation of
bound states of electrons at different sublattices.

The observability of the Mott insulating behavior of ANs depends critically
on the magnitude of the gap $\Delta _{\rho +}$ in the $\rho +$ mode, which is
estimated by the self-consistent harmonic approximation as \cite{power},
\begin{eqnarray}
& & \Delta_{\rho+}/(2 v_{\rho+}a^{-1}) \nonum \\
&=&
\left[\frac{K_{\rho+}}{\pi v_0} 
\sqrt{
\frac{\de V(0)^2}{2} + {\bar V}(2K_0)^2 - \de V(0) {\bar V}(2K_0)
} 
\right]^{\frac{1}{1-K_{\rho+}}} \hspace{-1cm}. 
\end{eqnarray}
Using the value of the matrix elements calculated numerically,
the charge gap for $N=10$ is estimated as 
$\sim 100 K$ for $a_0 = a/2$ and $\sim 10 K$ for $a_0 = a$
(in case of $a_0 = a$, 
$\de V(0) = 5.6 \times 10^{-3}$ and ${\bar V}(2K_0) = 6.9 \times 10^{-2}$ 
in units of $ae^2/2 \pi \kappa R$). 
The resistivity of ANs shows power law temperature dependence,  
$\rho \sim T^{2K_{\rho +}-1}/N^{2}$ 
at high temperatures $T \gg \Delta_{\rho+}$
and increases exponentially, $\rho \propto \exp(\Delta_{\rho+}/T)$,
at $T \ll \Delta _{\rho+}$ \cite{Kane-Balents-Fisher}. 
The temperature dependence of the resistivity at
half-filling is a characteristic signature of the Mott transition.
We conjecture that this signature can be best detected 
in multiprobe transport measurements \cite{Bezryadin}.

\strut

The authors would like to thank G. E. W. Bauer, R. Egger, and Yu. V. Nazarov
for stimulating discussions and L. Balents for critical reading of the
manuscript. 
The financial support of the Dutch Foundation
for Fundamental Research on Matter (FOM) is gratefully acknowledged. 
This work is also a part of INTAS-RFBR 95-1305. 


\begin{figure}[tbp]
\caption{ Solutions of the renormalization group equations for $K_{\rho+}$, $%
K_{\sigma+}$, $K_{\rho-}$ and $K_{\sigma-}$. The initial values of
parameters correspond to the estimates given in the text below Eq.(8) for $%
N=10$, $\kappa = 1.4$, $R_s = 100 nm$ and $a_0 = a/2$. Insert: The honeycomb
lattice of carbon atoms. Here $\vec a_{\pm}$ are the two primitive Bravais
lattice vectors, $|\vec a_{\pm}|=a$. The hexagon shown by thick line is the
unit cell and the black (white) circle denotes the point at $p=+(-)$
sublattice. The $x$ axis points along AN.}
\label{fig:1}
\end{figure}


\begin{references}
\bibitem{difference}  In fact, the kinetic term of Ref.\cite{Egger-Gogolin}
corresponds to the $y$-component of Eq. (\ref{eqn:1}) which would describe a
conducting zigzag nanotube parallel to the $y$ axis.
\bibitem{commentfbs}  Note that two-particle ''Umklapp'' scattering, $%
(\alpha ,\alpha )\leftrightarrow (-\alpha ,-\alpha )$, proposed in Ref. \cite
{Egger-Gogolin}, is forbidden due to violation of the quasi-momentum
conservation.
\end{references}
\end{document}